\documentclass{PoS}
\usepackage{graphicx}
\usepackage{amssymb}
\newcommand{\be}{\begin{eqnarray}}
\newcommand{\ee}{\end{eqnarray}}

\title{The GPDs of the photon}

\ShortTitle{The GPDs of the photon}

\author{\speaker{Asmita Mukherjee}\\
        Department of Physics, Indian Institute of Technology Bombay\\
        Powai, Mumbai 400076, India\\
        E-mail: \email{asmita@phy.iitb.ac.in}}

\abstract{We report on a recent calculation of the generalized parton
distributions (GPDs)  of the photon when the momentum transfer in the transverse
direction is non-zero. We use an overlap representation of the photon GPDs in
terms of the photon light-front wave functions. We calculate the GPDs at
leading order in electromagnetic coupling $\alpha$ and zeroth order in
strong coupling $\alpha_s$. We consider also the case when the helicity of
the photon is flipped. Fourier transform of the GPDs with respect to the
transverse momentum transfer gives the parton distributions of the photon in
impact parameter space. }

\FullConference{International Conference on the Structure and the
Interactions of
the Photon including the 20th International Workshop on Photon-Photon
Collisions and
the International Workshop on High Energy Photon Linear Colliders\\
                 20 - 24 May 2013\\
                 Paris, France}

\begin{document}
%%%%%%%%%%%%%%%%%%%%%%%%%%
\section{Introduction}
%%%%%%%%%%%%%%%%%%%%%%%%%%
The partonic substructure of the photon play an important role in high
energy scattering experiments where the virtuality of the photon involved 
is very large. The pointlike component of the photon structure function,
which is relevant in processes like  $e^+e^-$ annihilation and
photoproduction, can be calculated perturbatively. A striking feature of the
photon parton distribution is that they have logarithmic scale dependence
already at zeroth order of the strong coupling constant unlike the parton
distributions of the nucleon. The photon structure functions are now well
known both theoretically and experimentally. In \cite{pire} the authors have
considered deeply virtual Compton scattering on a photon target $\gamma^*
\gamma \to \gamma \gamma $ in the kinematical region of large center-of-mass
energy, large virtuality $(Q^2)$ but small squared momentum transfer $(-t)$
from the initial to final photon. They interpreted the results in terms of
the generalized parton distributions (GPDs)  of the photon at leading logarithmic
order, in analogy with the GPDs of the proton. At leading order in $\alpha$
and zeroth order in $\alpha_s$ the GPDs of the photon depend on the scale
logarithmically. These can be calculated perturbatively, unlike the proton
GPDs, and they can act as theoretical laboratories to understand the basic
properties of GPDs like polynomiality and positivity. In \cite{ter}  the GPDs of the
photon have been used to investigate the analyticity properties of DVCS
amplitudes and related sum rules of the GPDs. Recently we have investigated
the GPDs of the photon using overlaps of light-front wave functions (LFWFs) 
of the photon. We extended the study in \cite{pire} to a more general
kinematics when the momentum transfer between the initial and the final
photon also has a transverse component. We have shown that in this kinematics 
Fourier transform of the photon GPDs give impact parameter dependent parton
distribution of the photon in transverse position space. As is known in the
DVCS process involving a proton, the helicity of the proton may or may not
flip. When the helicity of the proton is flipped, the DVCS amplitude is
given in terms of the GPD $E$. The flip needs non-zero orbital angular
momentum (OAM) of the overlapping LFWFs of the proton, and is not possible
unless there is non-zero momentum transfer in the transverse direction. When
the nucleon is transversely polarized, this results in a distortion 
in the parton distributions in the impact parameter space. We have shown
that the photon GPDs where the helicity of the photon is flipped are related
to a similar distortion of the parton distributions of the photon in the transverse
impact parameter space which is due to the non-vanishing OAM of the partonic
constituents of the photon. 

%%%%%%%%%%%%%%%%%%%%%%%%%%%%%%%%%%%%%%%%%%%%%%%
\section{GPDs of the photon without helicity flip}
%%%%%%%%%%%%%%%%%%%%%%%%%%%%%%%%%%%%%%%%%%%%%%%%%%%

The GPDs for the photon can be written as the
following off-forward matrix
elements \cite{pire}:
   
\be
F^q=\int {dy^-\over 8 \pi} e^{-i P^+ y^-\over 2} \langle \gamma(P',\lambda') \mid
{\bar{\psi}} (0) \gamma^+ \psi(y^-) \mid \gamma (P,\lambda)\rangle ;
\nonumber\\ 
\tilde F^q=\int {dy^-\over 8 \pi} e^{-i P^+ y^-\over 2} \langle
\gamma(P',\lambda')
 \mid
{\bar{\psi}} (0) \gamma^+ \gamma^5 \psi(y^-) \mid \gamma (P,\lambda)\rangle .
\ee
$F^q$ contributes when the photon is unpolarized and $\tilde F^q$ is the
contribution from the polarized photon. We have chosen the light-front gauge
$A^+=0$ and used light-front coordinates. In this section, we take
$\lambda'=\lambda$ so that there is no helicity flip of the photon.  We use the Fock space expansion of 
a real photon of momentum $P$ and
helicity $\lambda$, which can be written as,
\be
\mid \gamma(P,\lambda)\rangle &=& \Big [ a^\dagger(P, \lambda) \mid 0
\rangle + \sum_{\sigma_1, \sigma_2} \int \{dk_1\} \int \{ dk_2\} \sqrt{2{(2
\pi)}^3 P^+} \delta^3 (P-k_1-k_2)\nonumber\\&&~~~~ \phi_2(k_1,k_2,\sigma_1,
\sigma_2)
b^\dagger(k_1, \sigma_1) d^\dagger(k_2, \sigma_2) \mid 0 \rangle \Big ]
\ee 
$\{ dk\}= \int {dk^+ d^2 k^\perp\over \sqrt{2 {(2 \Pi)}^3 k^+}}$, $\phi_2$ is the
two-particle ($q \bar{q}$) light-front wave function (LFWF) and $\sigma_1$
and
$\sigma_2$ are the helicities of the quark and antiquark. The wave function   
can be expressed in terms of Jacobi momenta $x_i={k_i^+\over P^+}$ and
$q_i^\perp=k_i^\perp-x_i P^\perp$. These obey the relations $\sum_i x_i=1,
\sum_i q_i^\perp=0$. The boost invariant photon LFWFs are given by     
${\psi_2(x_i,q_i^\perp)=\phi_2 \sqrt{P^+}}$. $\psi_2(x_i, q_i^\perp)$
can be calculated order by order in perturbation theory \cite{two}. The above
off-forward matrix elements can be calculated analytically using these
LFWFs \cite{ours1} and can be written as: 
\be
F^q &=& \sum_q {\alpha e_q^2 \over 4 {\pi}^2 } \Big [ ((1-x)^2+x^2) (I_1+I_2  
+L I_3) +2 m^2 I_3 \big ] \theta(x) \theta(1-x)  
\nonumber\\&&~~-\sum_q {\alpha e_q^2 \over 4 {\pi}^2 }
\Big [ ((1+x)^2+x^2) (I'_1+I'_2+ L'
I'_3) +2 m^2 I'_3 \Big ] \theta(-x) \theta(1+x)
\ee
Here the sum indicates sum over different quark flavors;
$L=-2 m^2 +2 m^2 x (1-x)
 - {(\Delta^\perp)}^2 (1-x)^2$,
$L'=-2 m^2 -2 m^2 x (1+x) - {(\Delta^\perp)}^2 (1+x)^2$; the integrals can be
written as,
\be
I_1=\int {d^2 q^\perp \over D} = \pi Log \Big [{\Lambda^2\over \mu^2
-m^2 x (1-x) +m^2}\Big ]=I_2\nonumber\\
I_3= \int {d^2 q^\perp\over D D'}= \int_0^1 d \alpha {\pi\over P(x,
 \alpha, {(\Delta^\perp)}^2)}
\ee
where $D={(q^\perp)}^2 -m^2 x (1-x) +m^2$ and $D'= {(q^\perp)}^2
+{(\Delta^\perp)}^2 (1-x)^2 -2 q^\perp \cdot \Delta^\perp (1-x) -m^2 x (1-x)   
+m^2$, and $P(x, \alpha, {(\Delta^\perp)}^2)= -m^2 x (1-x) +m^2 
+ \alpha(1-\alpha) (1-x)^2 {(\Delta^\perp)}^2$. At zeroth order in $\alpha_s$ the
results depend on the scale logarithmically. $\mu$ is a lower cutoff on the transverse 
momentum, which can be taken to zero as long as the quark mass is nonzero.

For the antiquark contributions we have similar integrals
\be
I'_1=\int {d^2 q^\perp \over H} = \pi Log \Big [{\Lambda^2\over \mu^2
+m^2 x(1+x)+m^2}\Big ]=I'_2\nonumber\\
I'_3= \int {d^2 q^\perp\over H H'}= \int_0^1 d \alpha {\pi\over Q(x, \alpha,
 {(\Delta^\perp)}^2)}
\ee
where $H={(q^\perp)}^2 +m^2 x (1+x) +m^2$ and $H'= {(q^\perp)}^2
+{(\Delta^\perp)}^2 (1+x)^2 +2 q^\perp \cdot \Delta^\perp (1+x) +m^2 x (1+x)
+m^2$, and $Q(x, \alpha, {(\Delta^\perp)}^2)= m^2 x (1+x) +m^2 
+ \alpha(1-\alpha) (1+x)^2 {(\Delta^\perp)}^2$.

For polarized photon the GPD $\tilde F^q$ can be calculated from the
terms of the form
$\epsilon^2_\lambda \epsilon^{1*}_\lambda-\epsilon^1_\lambda
\epsilon^{2*}_\lambda$ \cite{pire}. We consider the terms where the photon
helicity is not flipped. This can be written as,
\be
\tilde F^q &=& \sum_q {\alpha e_q^2 \over 4 {\pi}^2 } \Big [ (x^2-(1-x)^2)
(I_1+I_2+L
I_3) +2 m^2 I_3 \big ] \theta(x) \theta(1-x)
\nonumber\\&&+\sum_q {\alpha e_q^2 \over 4 {\pi}^2 } \Big [
(x^2-(1+x)^2) (I'_1+I'_2+ L'
I'_3) +2 m^2 I'_3 \Big ] \theta(-x) \theta(1+x)
\ee
In analogy with the impact parameter dependent parton distribution of the
proton \cite{burkardt}, we introduce the same for the photon. By
taking a Fourier transform with respect to the transverse momentum
transfer $\Delta^\perp$ we get the GPDs in the transverse impact
parameter space.
\be
q (x,b^\perp)&=&{1\over (2\pi)^2}\int d^2 \Delta^\perp
e^{-i\Delta^\perp \cdot b^\perp} F^q \nonumber \\
&=&{1\over 2 \pi}\int \Delta d\Delta J_0(\Delta b) F^q ;
\ee
\be
\tilde q (x,b^\perp)&=&{1\over (2\pi)^2}\int d^2 \Delta^\perp
e^{-i\Delta^\perp \cdot b^\perp} \tilde F^q \nonumber \\
&=&{1\over 2 \pi}\int \Delta d\Delta J_0(\Delta b) \tilde F^q ;
\ee
where $J_0(z)$ is the Bessel function; $\Delta=|\Delta^\perp|$ and
$b=|b^\perp|$.
In the numerical calculation, we have introduced a maximum limit
$\Delta_{max}$ of the $\Delta$
integration which we restrict to satisfy the kinematics $-t<<Q^2$
\cite{hadron_optics,chiral,quark,model}. $q(x,b^\perp)$ gives the
distribution
of partons in this case inside the photon in the transverse plane.
Like the proton, this interpretation holds in the
infinite momentum frame and there is no relativistic correction
to this identification because in light-front formalism,
as well as in the infinite momentum
frame, the transverse boosts act like non-relativistic Galilean boosts.
$q(x, b^\perp)$ gives simultaneous information about the longitudinal
momentum fraction $x$ and the transverse distance $b$ of the parton from the
center of the photon and thus gives a new insight to the internal structure
of the photon. The impact parameter distribution for a polarized photon is
given by $\tilde q(x, b^\perp)$. 

\begin{figure}[!htp]
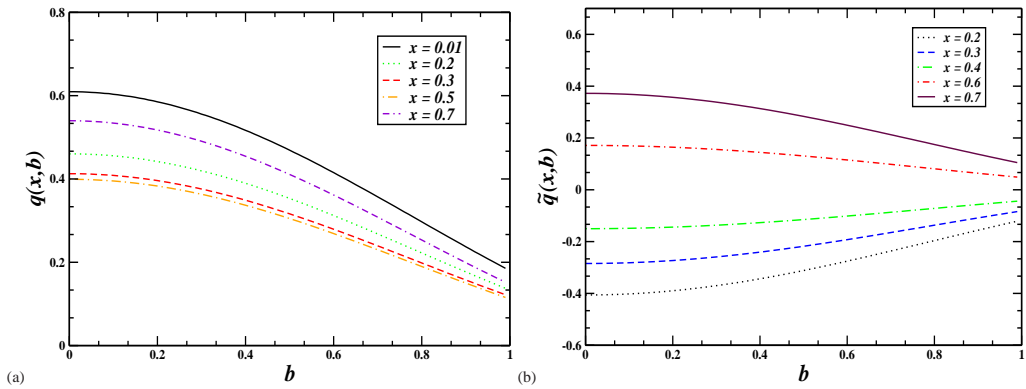

%\centering
\begin{minipage}[c]{0.9\textwidth}

\tiny{(a)}\includegraphics[width=6.5cm,height=5cm,clip]{fig2b.eps}
%\vspace{0.3in}
\tiny{(b)}\includegraphics[width=6.5cm,height=5cm,clip]{fig3b.eps}\\
\end{minipage}

\caption{\label{fig2}(Color online) Plots of impact parameter dependent
pdf $q(x,b)$ and $\tilde q(x,b)$ vs $b$ for fixed
values of $x$
where we have taken $\Lambda$ = 20 $\mathrm{GeV}$ and $\Delta_{max}$= 3
GeV where $\Delta_{max}$ is the upper limit in
the $\Delta$ integration. $b$ is in ${\mathrm{GeV}}^{-1}$ and  $q(x,b)$ and
$\tilde q(x,b)$  are in ${\mathrm{GeV}}^{2}$ .}
\end{figure}

Figs 1(a) and 1(b) show the impact parameter dependent parton distributions
$q(x,b)$ and $\tilde q(x,b)$ for unpolarized and polarized photons
respectively. We took the momentum transfer to be purely in the transverse
direction. We took the quark mass to be non-zero and equal to $m=m_q=3.3$
MeV; $\Lambda=Q=20 $ GeV; $\mu=0$ and $\Delta_{max}=3.0$ GeV, where $
\Delta_{max}$ is the upper limit of the Fourier transform. The smearing in
$b^\perp$ space reveals the partonic substructure of the photon and its
shape in transverse space. In the ideal definition of the Fourier transform
the integration over the $\Delta^\perp$ should be from zero to infinity. In
this case the $\Delta^\perp$ independent terms in the integrand
would give $\delta^2(b^\perp)$ in the  impact parameter space. This means
when there is no momentum transfer in the transverse direction, the photon 
behaves like a point particle in transverse position space. The distribution 
in transverse space  is revealed only when there is 
non-zero momentum transfer in the transverse direction. The parton
distributions for the polarized photon changes sign at $x=1/2$, at this
point the GPD and the ipdpdf become zero. These parton distributions for the
polarized photon are approximately symmetric about $x=1/2$ in $b$ space. For
fixed $x$, $\tilde q(x,b)$ becomes broader as a function of $b$ as $x$
increases till $x=1/2$. For larger values of $x$, it changes sign. We have
also checked that at larger values of $\Delta_{max}$ the distributions    
are sharper in $b$ space. The photon GPDs show qualitatively the same
behaviour when the momentum transfer has both longitudinal and transverse
components \cite{ours2}. 
    
%%%%%%%%%%%%%%%%%%%%%%%%%%%%%%%%%%%%%%%%%%%
\section{Helicity Flip Photon GPDs}
%%%%%%%%%%%%%%%%%%%%%%%%%%%%%%%%%%%%%%%%%%%%

In this case $\lambda' \ne \lambda$. Our calculations show that at leading
order there is only one helicity flip photon GPD. This can be expressed 
in terms of the two-particle LFWFs of the photon. As the photon has spin
one, to flip its helicity, one of the overlapping LFWFs should have orbital
angular momentum contribution of two units. 

The transverse polarization vector of the photon can be written as :
\be
\epsilon^\perp_\pm = \frac{1}{\sqrt{2}}(\mp1,-i)
\ee

We extract the GPD that involves a helicity
flip of the target photon from the non-vanishing coefficient of the
combination $(\epsilon^1_{+1} \epsilon^{1*}_{-1}+\epsilon^2_{+1} 
\epsilon^{2*}_{-1})$. Using the
overlap formula in terms of the LFWFs, the helicity flip photon GPD can be
calculated and has the form \cite{ours3} 
\be
E_1 = \frac{\alpha e_q^2}{2 \pi^2}x (1-x) \Big[ I_1 - (1-x)I_2 \Big].
\ee
The integrals are given by :
\be
I_1 =  ({(\Delta^1)}^2-{(\Delta^2)}^2)\pi (1-x)^{2}
\int_0^{1} dq\hspace{0.2cm}
\frac{(1-q)^2}{B(q)};~~~   
I_2= ({(\Delta^1)}^2-{(\Delta^2)}^2) \pi (1-x) \int_0^{1} dq\hspace{0.2cm}
\frac{(1-q)}{B(q)}; 
\ee
where  
\be    
B(q) = m^2\Big(1-x (1-x)\Big) +q (1-q) (1-x)^2
{(\Delta^\perp)}^2.
\ee
$\Delta^1$ and $\Delta^2$ are the $x$ and $y$ components of $\Delta^\perp$
respectively. 
   
%%%%%%%%%%%%%%%%%%%%%%%%%%%%%%%%%%%%%%%%%%%%%%%%%%%%%%%%%%%%%%%%%%%%%%%%%%%%%%%%%%%%%%%5555
\begin{figure}
\begin{minipage}[c]{0.99\textwidth}
\tiny{(a)}\includegraphics[width=7cm,height=6cm,clip]{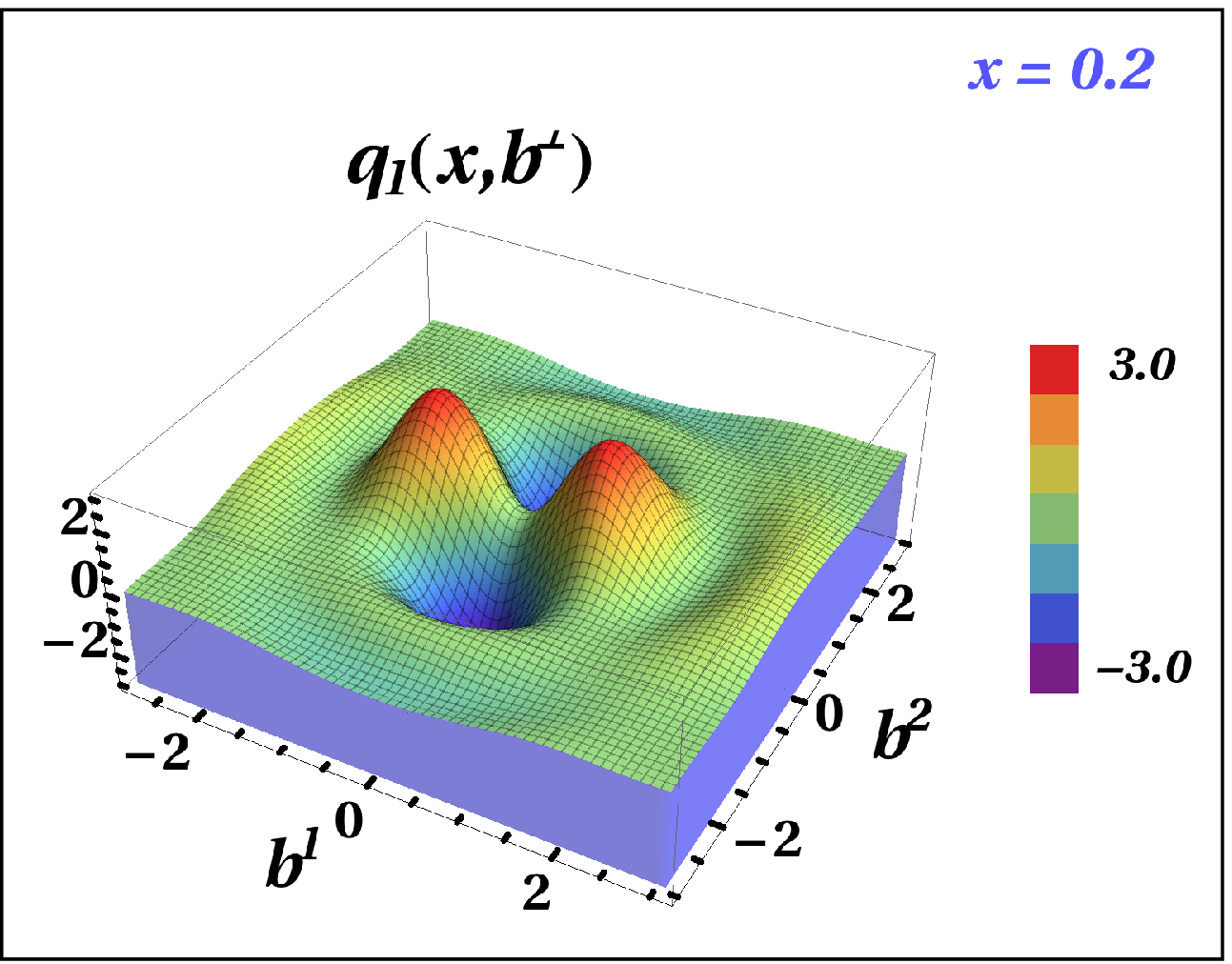}
\hspace{0.1cm}
\tiny{(b)}\includegraphics[width=7cm,height=6cm,clip]{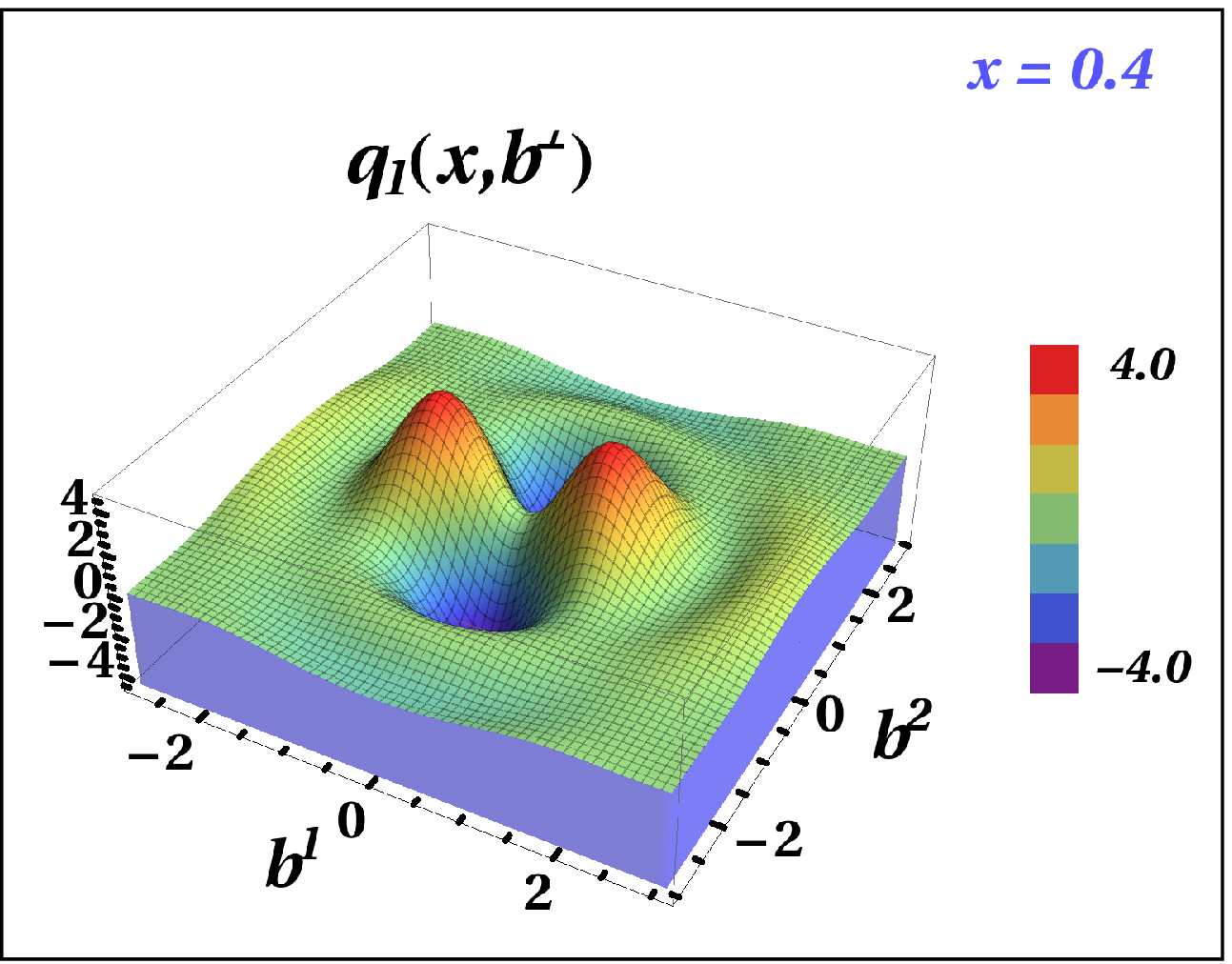}
\end{minipage}
\caption{\label{fig5}(Color online) Plots of $q_1 (x,b^\perp)$ vs $b^1,b^2$
for
different values of $x$ and at $\Delta_{max} = 4.0$ GeV.}
\end{figure}
%%%%%%%%%%%%%%%%%%%%%%%%%%%%%%%%%%%%%%%%%%%%%%%%%%%%%%%%%%%%%%%%%%%%%%%%%%%%%%

\vspace{0.2in}

Similar to the GPD $E$ of a spin $1/2$ particle for example  a dressed
electron/quark
\cite{quark}, the helicity flip photon GPD has no
logarithmic scale dependent term.  
From the expressions above, we define the 
parton distributions \cite{burkardt} with  the helicity flip of the photon
in transverse impact  parameter space as:
\be
q_1(x,b^\perp)={1\over 4 \pi^2} \int d^2 \Delta^\perp e^{-i \Delta^\perp
\cdot b^\perp} E_1 (x,\Delta^\perp);
\ee
where $t=-{(\Delta^\perp)}^2$ and $b^\perp$ is the transverse impact
parameter conjugate to $\Delta^\perp$. One obtains :
\be
q_1(x,b^\perp)=\frac{1}{4\pi^2} \int d^2\Delta^\perp e^{-ib^\perp \cdot
\Delta^\perp} ({(\Delta^1)}^2-{(\Delta^2)}^2) f(x) Q(x,t),
\ee 
where
\be  
f(x)= \frac{\alpha e_{q}^2}{2 \pi}x (1-x)^{3}
\hspace{0.5cm}, ~~~~ \hspace{0.5cm} Q(x,t)=  
\int_0^{1} \frac{dq}{B(q)}\hspace{0.2cm}     
((1-q)^2-(1-q));
\ee 
Figs 2(a) and 2(b) show plots of the helicity flip photon GPDs in impact
parameter space $q_1(x,b)$ vs $b_1$ and $b_2$ for a fixed value of
$\Delta_{max}$ and two different values of $x$. The magnitude of the peak
depends on $x$. $q_1(x,b)$ has a quadrupole structure, as can be seen in the
plots. This structure is expected as to flip the helicity of a spin one 
object, one needs a LFWF with orbital angular momentum of two units.     

%%%%%%%%%%%%%%%%%%%%%%%%%%%%%%
\section{Conclusion}
%%%%%%%%%%%%%%%%%%%%%%%%%%%%%
In this talk, I have reported a recent calculation of the generalized parton
distributions of the photon. Extending the calculations of \cite{pire},
from the kinematics of zero $\Delta^\perp$ and non-zero $\zeta$, we
calculated the photon GPDs for non-zero $\Delta^\perp$, at leading order in
$\alpha$ and zeroth order in $\alpha_s$. The GPDs where the helicity of the
photon is not flipped are logarithmically dependent on the scale. Taking the
Fourier transform of the GPDs with respect to $\Delta^\perp$ we get the
parton distributions of the photon in impact parameter space. The helicity
flip photon GPDs show the expected quadrupole structure in impact parameter
space.  

%%%%%%%%%%%%%%%%%%%%%%%%%%%%%%
\section{Acknowledgement}
%%%%%%%%%%%%%%%%%%%%%%%%%%%%%

This work has been done in collaboration with Sreeraj Nair and Vikash Kumar
Ojha. AM thanks the organizers of the 
"International Conference on the Structure and the Interactions of the
Photon" for the kind invitation.

\end{document}